\documentclass[aps,prd,preprintnumbers,nofootinbib,twocolumn]{revtex4}
\usepackage{bm}
\usepackage{latexsym}
\usepackage{dcolumn}
\usepackage{amsmath,amsfonts,amssymb}
\usepackage{graphicx,epsfig}
\usepackage{psfrag}
\usepackage{amsthm}
\usepackage{color}

\usepackage{bm}
\usepackage{latexsym}
\usepackage{dcolumn}
\usepackage{amsmath,amsfonts,amssymb}
\usepackage{graphicx,epsfig}
\usepackage{psfrag}
\usepackage{amsthm}
\usepackage{color}

\usepackage{nccmath}
\usepackage{moresize}
\usepackage{enumerate}
\usepackage[hang,flushmargin]{footmisc}
\usepackage[titletoc,toc]{appendix}
\usepackage{lipsum}
\usepackage{subcaption}
\usepackage{epstopdf} 
\usepackage{hyperref}
\usepackage{rotating}
\usepackage{multirow}
\usepackage{mathtools}

\usepackage{stackengine}
\usepackage{scalerel}

\newcommand{\be}{\begin{equation}}
\newcommand{\ee}{\end{equation}}
\newcommand{\bea}{\begin{eqnarray}}
\newcommand{\eea}{\end{eqnarray}}
\newcommand{\bse}{\begin{subequations}}
\newcommand{\ese}{\end{subequations}}
\newcommand{\bce}{\begin{center}}
\newcommand{\ece}{\end{center}}
\newcommand{\bfg}{\begin{figure}}
\newcommand{\efg}{\end{figure}}
\newcommand{\bit}{\begin{itemize}}
\newcommand{\eit}{\end{itemize}}
\newcommand{\bed}{\begin{description}}
\newcommand{\eed}{\end{description}}
\newcommand{\ben}{\begin{enumerate}}
\newcommand{\een}{\end{enumerate}}
\newcommand{\nn}{\nonumber}

\newcommand{\pa}{\partial}
\newcommand{\fr}{\frac}

\newcommand{\no}{\noindent}

\def\le {\left}
\def\ri {\right}
\def\m  {\mu}
\def\n  {\nu}

\def\r  {\rho}

\newcommand{\cR}{\mathcal R}
\newcommand{\cS}{\mathcal S}

\newcommand{\vx}{\vec{\pmb x}}

\newcommand{\vJ}{\vec{\pmb J}}

\newcommand{\bdm}{\begin{displaymath}}
\newcommand{\edm}{\end{displaymath}}

\begin{document}

\title{Emergent gravity and the quantum}

\author{Saurya Das} \email[email: ]{saurya.das@uleth.ca}
%


\affiliation{Theoretical Physics Group,
Department of Physics and Astronomy,
University of Lethbridge, 4401 University Drive,
Lethbridge, Alberta T1K 3M4, Canada}

\author{Sourav Sur}
\email[email: ]{sourav.sur@gmail.com}

\affiliation{Department of Physics and Astrophysics, University of Delhi, New Delhi 110007, India}

\begin{abstract}
We show that if one starts with a  
Universe with some matter and a cosmological constant, then quantum mechanics naturally induces an attractive gravitational potential and an effective Newton's coupling. Thus gravity is an emergent phenomenon and what should be quantized are the fundamental degrees of freedom from which it emerges.

\vspace{0.3cm}
\noindent
%
{\bf This essay received an Honorable Mention in the 2021 Gravity Research Foundation Essay Competition}\\

\end{abstract}

\maketitle

Despite decades of intensive research, 
a theory of quantum gravity remains as elusive as ever.
It is well-known that perturbative quantum gravity, with metric fluctuations $h_{\m\n}$ as the fundamental degrees of freedom, is non-renormalizable \cite{feynman}. 
While on the one hand, this has inspired many 
candidate theories of quantum gravity \cite{string,lqg}, 
on the other, it has led to 
the suggestion that gravity may be an emergent phenomenon and the metric fluctuations may not be the right degrees of freedom to quantize after all \cite{jacobson,emergent}.

In this essay, we propose a specific model of emergent gravity based on a very few assumptions. 
In particular, we show that if one starts with
a given matter density $\r$ (say a dark matter density) and a cosmological constant $\Lambda$, then gravity
emerges naturally via quantum mechanics. 
That is, gravity can be a viewed as a
derived interaction, with the burden of quantization shifting to its more fundamental precursors. 
To see this, let us first review a lesser known, but perfectly legitimate version of quantum mechanics, 
in which a wavefunction written in the form 
\be \label{wf1}
\Psi (\vx,t) =\, \cR (\vx, t) \, e^{i \,\cS(\vx,t)} \, \quad \le[\cR, \cS \in \mathbb{R}\ri] \,, 
\ee 
when substituted in a relevant wave equation, e.g. the Schr\"odinger equation, yields two real equations as follows \cite{bohm}:
\bea 
&& \vec \nabla \cdot \vJ +\, \fr{\pa\r}{\pa t} =\, 0 \,, \label{consv-eq} \\
&& m \, \fr{d\vec v}{dt} =\, - \, \vec\nabla \le(V +\, V_Q\ri) \,,  \label{Newt-eq}
\eea 
where the `velocity field' of constituent particles is defined as
$\vec v \equiv (\hbar/m)\,\vec\nabla S$. 
While Eq.\,(\ref{consv-eq}) is just the standard conservation equation of the probability density
and current ($\r$ and $\vJ$ respectively), Eq.\,(\ref{Newt-eq}) shows, quite remarkably, that 
the quantum dynamics of a particle of mass $m$ can be predicted from a classical equation (Newton's 
law in this case), albeit with the classical potential $V$ augmented by a {\em quantum potential} 
\be \label{QP}
V_Q =\, -\, \fr{\hbar^2}{2m} \fr{\nabla^2 \cR} \cR \,. 
\ee 
Note that $V_Q$ depends on the wavefunction. 
For instance, while a stream of free electrons, governed just by a vanishing classical potential 
($V = 0$), follows straight line trajectories, that governed by $(V + V_Q)$ for an appropriate 
choice of $\psi(\vx,t)$ reproduces the familiar interference patterns and dark and light fringes 
\cite{fringes}.

Now, let us come to 
an even lesser known result which follows directly from 
Eqs.(\ref{Newt-eq}) and \,(\ref{QP}), 
namely that for a {\em stationary} state, with energy $E$ and described by the wavefunction
$\, \Psi (\vx,t) = \psi(\vx)\, e^{-i E t/\hbar}\,$, one 
can easily show that 
\footnote{
%
The easiest way is to see this is to
recognize that 
$\nabla^2 {\cal R}/{\cal R} = \nabla^2 {\psi}/{\psi}$
for stationary states, and
substituting for $\nabla^2 \psi$ using the time-independent
Schr\"odinger equation 
%
$-\, \fr{\hbar^2}{2m}\, \nabla^2 \psi +\, V \psi =\, E \psi \, $
%
%
in Eq.(\ref{QP}) to obtain Eq.(\ref{QP1}).
}
\be \label{QP1}
V_Q =
\, -\, V +\, E \, . 
\ee
The result is remarkable for two reasons. First, it shows that the $\hbar$'s cancel out from Eq.(\ref{QP}) and the induced quantum potential is simply equal and opposite to the starting classical potential
$V$ (up to the unimportant constant $E$). Accordingly, the quantum force is also equal and opposite to the starting classical force. 
%
Second, writing the above as
\be \label{QP2}
V =\, -\, V_Q +\, E \, ,
\ee
it follows that if we had started with $V_Q$ instead of $V$ in the Schr\"odinger equation, it would 
have induced an equal and opposite $V$ via quantum mechanics! 
%
In other words, what is to be treated as classical and what as quantum is a matter of choice, and the presence of {\em any} one would 
cause the other to emerge naturally, 
and this is true regardless of the exact form of the starting potential.

This has an interesting implication in cosmology, as we show below. 
We present the main points here and the details can be found in 
\cite{sd,db1,db2,db3,dss}.
Consider the standard spatially flat 
Universe, with the 
critical (or, total) density $\r_{_C}$, and
having a given dark matter density $\r\simeq \r_c$ 
(e.g. $\r=0.25\,\r_c$ in the present epoch).
The classical (Newtonian) gravitational interaction experienced by a DM particle of mass 
$m$ due to the background density, 
is given by 
\footnote{One can think of the mass $m$ on the surface of a sphere of radius $r$. Then only the mass inside the sphere contributes to the gravitational forces on it, given by Eq.(\ref{newton3}).}
\bea 
m \, \ddot{\vx} &=& -\, \fr{G m M}{|\vx|^2} \,\hat{x}-\, \vec\nabla V_Q \,\,, \nn 
\\
&=& -\fr{G m}{|\vx|^2} \le(\fr{4\pi} 3 \, |\vx|^3 
\rho \ri) \,\hat{x}
-\, \vec\nabla V_Q \,\,, 
\label{newton3}
\eea
with $\vx$ denoting the radius vector from any point in the comoving 
frame, $\hat x = \vx/|\vec x|$, 
dot $\{\cdot\} \equiv d/dt\,$ (with $t$ being the comoving time), and we 
have included the quantum potential term. 
Writing the spatial distance function $|\vx| \propto 
a(t)$, the scale factor of our Universe, we get 
\begin{eqnarray}
&& \frac{\ddot a}{a} 
= - \frac{4\pi G}{3}\,\left( \rho +3p \right) 
+ \frac{\Lambda}{3}~,
\label{fe1}
\end{eqnarray}
where $\Lambda = - \nabla^2 V_Q $, we again used $\rho \simeq \rho_c$,
and the pressure term is included for completeness. 
Next, one assumes that dark matter consists of a Bose-Einstein condensate of light bosons of mass $m$, stretching across cosmological length scales and described by 
the following macroscopic wavefunction
%
\begin{eqnarray}
\Psi (\vx,t) = \frac{R_0}{a^{3/2}}\, e^{-|\vx|^2/\sigma^2}\ e^{- i E_0 t/\hbar}~,
\label{psi1}
\end{eqnarray}
where $\sigma^2 = 2\sqrt{2}\,\hbar/(m H_0)$
and $E_0$ is the ground state energy of the condensate.  
Note that the above 
wavefunction (\ref{psi1}) is not only 
consistent with the large scale isotropy of the universe, but also ensures the dilution of the 
dark matter density with time ($\r = |\Psi|^2 \sim a^{-3}$), at least for 
time-scales $t \ll H_0^{-1}$, 
where $H_0$ is the Hubble parameter \cite{db1,db2,db3,dss}. 
It was shown in \cite{db1} that one needs
$m\leq 6\,eV/c^2$, such that the critical temperature of the Bose-Einstein condensate exceeds the ambient temperature of the Universe for all epochs, and the condensate can form in the early Universe. 
%
For the above $\Psi$, one can compute the $V_Q$ and $\Lambda$ and one  
finally gets
$\Lambda  \simeq H_0^2$.  
This implies the corresponding density 
$\rho_\Lambda =\Lambda/4\pi G 
\simeq \rho
$ and the pressure
$p_\Lambda = - \rho_\Lambda$, where we have used
$\rho_c = 3H_0^2/8\pi G$.
That is, one naturally arrives at the 
result that $\Lambda >0$ and 
$\rho_\Lambda \simeq \rho$,
just as observed currently in nature! 
One does not need to extremely
fine-tune $\Lambda$ or
$\rho$ in the early Universe to explain their approximate equality 
at the present epoch, known as the coincidence problem. There is no coincidence problem in this picture \cite{dascoincidence}. 

Having shown that the classical gravitational interaction can induce a 
cosmological constant via quantum mechanics, we now
go one step further, and essentially reverse the set of steps
from Eq.(\ref{fe1}) to Eq.(\ref{newton3}). In other words, we imagine a
Universe which is spatially flat, and endowed with a certain energy density $\rho \simeq \rho_c$ and a cosmological constant 
$\Lambda$. Then for a stationary state of the matter, one would get, in lieu of Eq.(\ref{newton3})
\begin{eqnarray}
m \ddot \vx = -\frac{\Lambda\,Mm}{8\pi \rho_c\,|\vx|^2}  
- \vec\nabla\, \nabla^{-2} \left( -\Lambda\right) ~. 
\label{reversenewton}
\end{eqnarray}
The second term now corresponds to the repulsive potential associated with the cosmological constant, 
while the first is the induced gravitational force, with the
effective induced Newton's coupling that can be read-off from Eq.(\ref{reversenewton}) as
\begin{eqnarray}
G = \frac{\Lambda}{8\pi \rho_c}~.
\label{gfroml}
\end{eqnarray}
Eq.(\ref{gfroml}) above summarizes our main result.
It shows that gravity and 
Newton's constant emerges naturally, as a {\it quantum phenomenon}
from a {\it classical} cosmological constant and matter density.  
Therefore if one wishes to quantize this system, say as a field theory, 
one would need to work with the degrees of freedom that describe the cosmological constant and its dynamics. For example, the cosmological constant may be taken to be
of the form $\Lambda = \Lambda_0 + \delta \Lambda$, where $\Lambda_0=$ constant and 
the dynamics of $\delta \Lambda$ is governed by a quantum field theory, such as that of a potential energy dominated scalar field. 
One just needs a sufficient abundance of light bosons and a $\Lambda$ to go with. 
The issue of renormalizability of the system should be seen from this perspective and the predictions can of course be translated to effective gravitational interactions  
by the prescription given earlier. 
It would be interesting to study this. 

%


\noindent 
{\bf Acknowledgment}

\no
%
This work was supported by the Natural Sciences and Engineering
Research Council of Canada.
%



\end{document}